\documentclass[namedreferences]{solarphysics}
\usepackage[hyperref,optionalrh,solaromanenum]{spr-sola-addons} 
\usepackage{graphicx}                    
\usepackage{amssymb}                    
\usepackage{color,xcolor}                       
\usepackage{soul}
\usepackage{ulem}
\soulregister\cite7
\soulregister\citealp7
\soulregister\ref7

\chardef\us=`\_

\begin{document}

\begin{article}
\begin{opening}

\title{Characterization and Correction of the Scattering
Background Produced by Dust on the Objective Lens of
the Lijiang 10-cm Coronagraph
}

\author[addressref={ad1,ad2,ad3},email={shafeiyang@ynao.ac.cn}]{\fnm{Feiyang }\lnm{Sha}}
\author[addressref={ad1,ad2,ad3},corref,email={lyu@swjtu.edu.cn}]{\fnm{Yu }\lnm{Liu}}
\author[addressref={ad1},corref,email={zhangxuefei@ynao.ac.cn}]{\fnm{Xuefei }\lnm{Zhang}}
\author[addressref={ad1},email={stf@ynao.ac.cn}]{\fnm{Tengfei }\lnm{Song}}

%
\runningauthor{Sha, F., et al.}
\runningtitle{Scattering Background by Dust on a Coronagraph Objective Lens}

\address[id={ad1}]{Yunnan Observatories, Chinese Academy of Sciences, Kunming, China}
\address[id={ad2}]{School of Physical Science and Technology, Southwest Jiaotong University, Chengdu, China}
\address[id={ad3}]{School of Astronomy and Space Science, University of Chinese Academy of Sciences, Beijing, China}

\begin{abstract}
Scattered light from the objective lens, directly exposed to the intense sunlight, is a dominant source of stray light in internally occulted coronagraphs. 
The variable stray light, such as the scatter from dust on the objective lens, can produce varying scattering backgrounds in coronal images, significantly impacting image quality and data analysis. 
Using data acquired by the \textit{Lijiang 10-cm Coronagraph}, the quantitative relationship between the distribution of dust on the objective lens and the resulting scattering backgrounds background is analyzed. 
Two empirical models for the scattering background are derived, and used to correct the raw coronal data. The second model, which depends on three parameters and performs better, shows that the scattering-background distribution varies with angle, weakens with increasing height, and enhances with increasing dust level on the objective lens. Moreover, we find that the dust on the center of the objective lens can contribute more significantly to the scattering background than on the edge. 
This study not only quantitatively confirms the significant impact of the stray light produced by dust on the objective lens of the coronagraph,  but also corrects the coronal data with this stray light for the first time.
Correcting for dust-scattered light is crucial for the high-precision calibration of ground-based coronagraph data, enabling a more accurate analysis of coronal structures. Furthermore, our model is envisioned to support the provision of reliable observational data for future routine coronal magnetic-field measurements using ground-based coronagraphs.

\end{abstract}
%
\end{opening}

%
\section{Introduction}
	\label{Intro} 

The corona is the outermost layer of the solar atmosphere, containing a thin, highly ionized plasma that reaches a temperature of a few million degrees. It has an extremely low brightness and diminishes quickly with height, i.e. radial distance from the solar center. Therefore, the corona is hidden by the much brighter solar disk and is generally not visible in normal conditions (e.g. \citealp{Habbal10,Liang21}).
The corona can be observed if the disk light is occulted, e.g. during a total solar eclipse or using a Lyot coronagraph (\citealp{Lyot33}).

Around the beginning of the 20th century, many attempts had been made to observe the solar corona during noneclipse periods. The most difficult part of coronal observation was suppressing the relatively strong stray light, which is typically two orders of magnitude higher than coronal brightness (\citealp{aliev17}). It was not until 1930 that \cite{Lyot33} directly observed the solar corona during noneclipse periods at the Pic du Midi Observatory at a latitude of 2780 m.  
The instrument he used is now known as the Lyot coronagraph, which employs internal occultation. The basic principle of this instrument involves placing an occulting disk inside the telescope to block the solar disk formed by the objective lens and enable the corona light to pass through. However, it requires strict suppression of stray light. Stray light can originate from various sources inside the instrument (\citealp{lyot39}):
\begin{enumerate}
	\item Direct sunlight from the solar photosphere.
	\item The ghost image formed by multiple reflections of the Sun on the two faces of the objective lens.
	\item The stray light from direct sunlight striking the telescope tube and the edge of the objective lens.
	\item The scattered light produced by surface roughness, scratches, and internal bubbles of the objective lens under direct sunlight.
	\item The stray light produced by dust particles on the objective lens.
\end{enumerate}
The conjugate-blocking method is used to suppress the first three types of stray light by positioning a stop at the conjugate image of the source.
To minimize stray light resulting from surface roughness and defects on the objective lens, high-quality materials and ultrasmooth polishing are utilized in lens construction. Regular lens cleaning is also crucial to suppress stray light originating from dust particles on the objective lens. 

The stray light outside the instrument primarily originates from atmospheric scattering, known as sky brightness.
Therefore, ideal coronal-observation sites should be located at high altitudes with low sky brightness. Additionally, stable and gentle winds, high levels of sunshine and a high number of sunny days are also necessary. 

It is crucial to understand that there are two types of stray light: fixed and variable. 
Fixed stray light remains constant, while variable stray light increases rapidly with time and deterioration of cleanliness level.
For routine ground-based coronagraph observations, the variable stray light is contributed by sky brightness and dust on the objective lens. 

Therefore, coronal data observed at different times can exhibit varying levels of scattered stray light, leading to different degrees of scattering background in coronal images. This makes it challenging to analyze subtle coronal structures and calibrate coronal intensity accurately. In addition, many methods have been proposed to measure the coronal magnetic field using coronagraphs (e.g., \citealp{Yang20Sci,Chen23RAA,Chen23MN}), which cannot be achieved without accurate data calibration.

Therefore, there is an urgent need for better methods to remove the scattering background. One simple way is to clean the objective frequently or use gas purging. However, this method has many drawbacks. First, dust accumulation can still occur and produce variable stray light, because of the atmospheric particles or low cleanness level, which is beyond our control. Secondly, there is no standard on whether the objective should be cleaned or not. Hence, cleaning is quite subjective, leading to over- or undercleaning. Finally, frequent cleaning of the objective lens can damage the lens itself, causing extra stray light.

Another possible way is image correction, which requires modeling the scattered light from dust on the objective lens. \cite{paul} measured the scattered light from dust-contaminated mirrors at $\lambda=10.6\ \rm\mu m$ and the results were compared to those predicted by a modified Mie theory. However, this study applies only to reflective optics and not to the refractive optics used in most ground-based coronagraphs. Based on Mie theory, \cite{Dittman} simulates the Bidirectional Scattering Distribution Function (BSDF) of scatter from particulate contaminated optics, where the forward and backward scatter BSDF are separated, allowing the analyst to consider these scatter contributions separately for refractive systems. \cite{Gallagher} use the Harvey Shack BSDF model to simulate the scattering distribution of dust on the objective lens of the \textit{Coronal Solar Magnetism Observatory Large Coronagraph} (COSMO-LC). However, these studies are based on theoretical simulations and few empirical measurements.

Based on the experimental data of \textit{Lijiang 10-cm Coronagraph}, \cite{sha23} derived an empirical model for the relation between the dust on the objective lens and the resulting scattering background on the coronal images. However, the employed strategy does not take into account the spatial distribution of dust particles and a one-dimensional model that is only related to height is obtained.

In this study, we propose a new dust-scattering background empirical model based on the experimental data of \textit{Lijiang 10-cm Coronagraph} (\citealp{ichimoto, 2018IAUS..340..169L}; \citealp{Liu14}).
\section{Experiment}
\label{exp}
\subsection{Experimental Conditions and Instruments	}
We conducted the experiment at Lijiang Observatory (3200\ m, E: $100^\circ01^\prime48^{\prime\prime}$, N: $26^\circ41^{\prime}42^{\prime\prime}$), the first observation site in China capable of coronal observations, with an average sky brightness of less than 20 ppm of the solar-disk intensity (\citealp{zhao18}). The instrument used is the \textit{Lijiang Coronagraph} (YOGIS), the only routinely operated coronagraph in China. It has been in continuous operation since it was established at Lijiang Observatory in 2013. 

The observation setup consists of a 10-cm coronagraph, a tunable Lyot filter, and a cooled CMOS camera, allowing it to image the coronal green line (Fe XIV, 530.3 nm, \citealp{Sakurai99}) within 1.03 to 2.5$\rm\  R_\odot$ field of view. The transmission curve of the Lyot filter can be modulated via two liquid-crystal variable readers, providing quick wavelength tuning and an efficient subtraction of sky background (\citealp{ichimoto}). 

In addition, YOGIS has the ability to obtain 2D Doppler velocity maps of the corona from which \cite{Sakurai02} found slow sound waves in the corona and \cite{Hori} detected an MDH kink oscillation in coronal loops for the first time. 
Using this instrument, \cite{Zhang_sdo} found that the coefficients between the green-line coronal data and the 211 Å data of the Atmospheric Imaging Assembly (AIA) onboard the Solar Dynamics Observatory (SDO) (\citealp{aia}) always kept high values. \cite{Zhang_mag} also suggested that the correlation between the coronal green line and the magnetic field was strong. \cite{Li23} conducted an analysis of intensity attenuation of the coronal structures and loops within the inner corona region. The above-mentioned study was affected by the dust-scattering background overlapped with the coronal data. If this background is removed, more accurate results can be obtained.

We seek to determine the relationship between dust on the objective lens and the resulting scattering background of the coronal image. Therefore, it is crucial to first obtain their respective individual contributions.

\subsection{Scattering Background}
\begin{figure}
	\centering
	\includegraphics[width=1\linewidth]{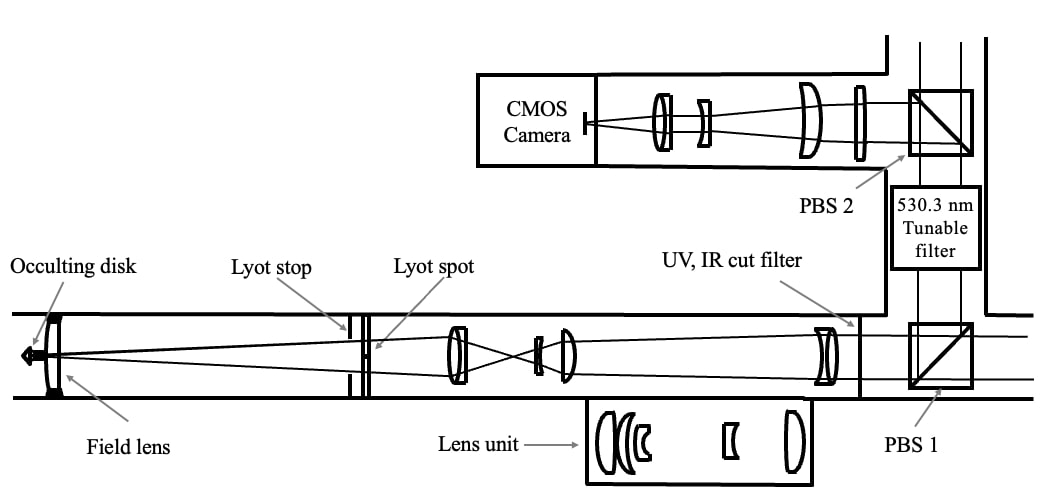}
	\caption{The optical layout of YOGIS. The objective lens, with a diameter of 10 cm and a focal length of 1490 mm are not depicted in the diagram. The occulting disk, on the far left of the figure, is positioned at the primary focus of the objective lens and it is attached at the center of the field lens. The Lyot stop is used to suppress the stray light originating from the telescope tube and the edges of the objective lens, while the Lyot spot suppresses the ghost image. Two polarizing beamsplitter (PBS1 and PBS2) are placed before and after the Lyot filter, not only splitting the beam but also working as the entrance and exit polarizers of the Lyot filter (\citealp{ichimoto}). The lens unit, not included in the optical path, is used for observing the objective lens.
	}
	\label{corona}
\end{figure}
The dust on the objective adds a scattering background to the coronal image.
This scattering background can be derived by removing the pure coronal part of the image, which can be approximately obtained after wiping out the dust on the objective. 

The optical layout of YOGIS (\citealp{sha23}) is shown in Figure~\ref{corona}. 
When imaging the corona, the filter is first adjusted to 530.3 nm with a bandwidth of 0.1 nm, corresponding to the Fe {\footnotesize XIV} line center, to obtain a ``single" image. Then, the filter is adjusted to the line wing (530.3 $\pm$ 0.2 nm) to obtain a ``Double" image for removing the sky background. The coronal image is obtained using the following equation:
\begin{equation}
	\mbox{Corona}=\frac{\mbox{Single}-\mbox{Double}}{\mbox{Flat}-\mbox{Dark}}\times\overline{\mbox{Flat}-\mbox{Dark}}.
\end{equation}

To obtain the scattering background, the corona is imaged with dust on the objective first, denoted as $C_0$. Then, the lens is cleaned three times to remove the dust, with $C_1, C_2$, and $C_3$ denoting the images obtained after each cleaning, respectively. We define three scattering-background images:
\begin{equation}
	\mbox{Scatter}_i=C_i/s_i-C_3/s_3\ \ (i=0,1,2),
	\label{c/s}
\end{equation}
where $s_i$ is the solar-radiation intensity and is used to correct for its diurnal variation. These coronal images are aligned in advance.

\subsection{Dust on the Objective Lens}
\begin{figure}
	\centering
	\includegraphics[width=1\linewidth]{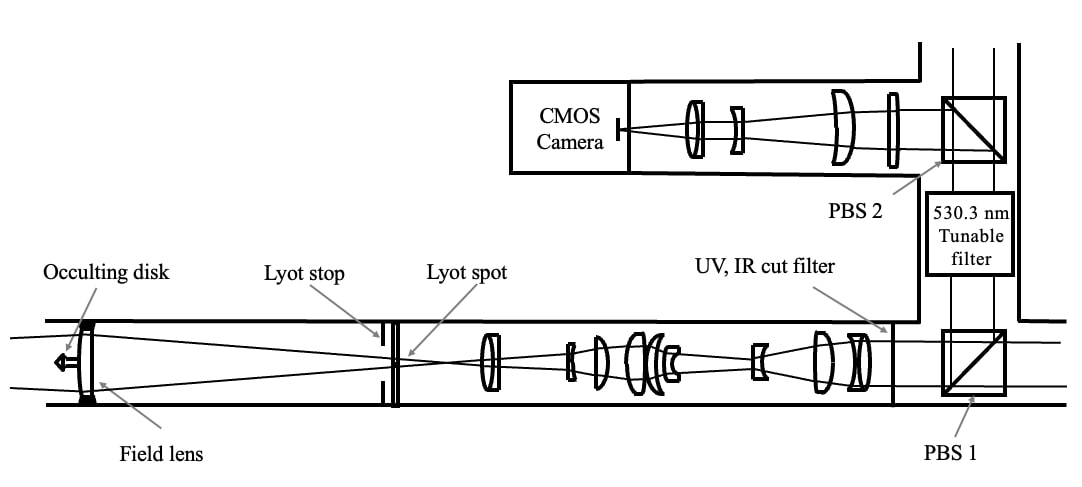}
	\caption{The optical layout of YOGIS when imaging the objective lens. The lens unit is moved in the optical path compared to Figure~\ref{corona}. It serves to extend the optical path, bringing the original position of the objective lens image forward to the location of the camera.}
	\label{lens}
\end{figure}

Inside YOGIS, there is a lens unit mounted on a one-dimensional linear stage, which enables making an image of the objective lens on the camera to evaluate the scattered light produced by dust (\citealp{ichimoto}). The optical layout, when imaging the objective is shown in Figure~\ref{lens}.

\begin{figure}
	\centering
	\includegraphics[width=.7\linewidth]{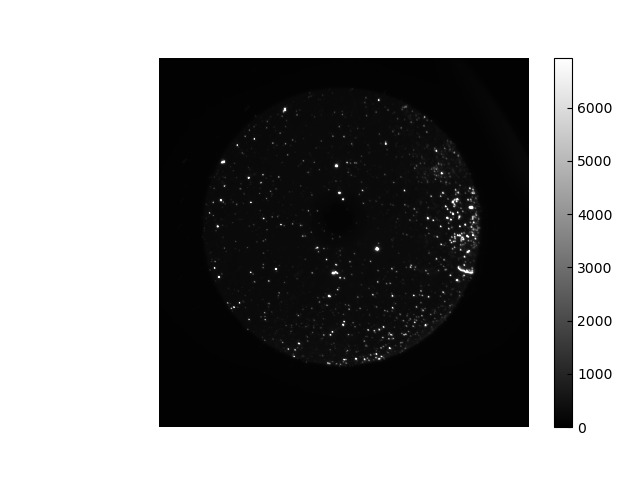}
	\caption{Image of the objective in arbitrary units. The large circle in the image represents the image of the objective lens, resulting from the scattered light produced by the surface roughness of the objective lens. The numerous small spots are scattering points formed by dust particles on the objective lens surface. }
	\label{img_lens}
\end{figure}

The image of the objective lens, shown in Figure~\ref{img_lens}, is produced by three sources. First, the sunlight scattered by the surface roughness of the objective lens. Secondly, the sunlight scattered by the dust on the objective lens, forming points of different sizes with approximately circular shapes. The third source is the scattered light from Earth's atmosphere, scattered by dust on the objective lens.
Among them, the first source only depends on the solar-radiation intensity and the objective lens surface-processing technology, which is fixed with respect to the change in environmental cleanliness level. In the third source, the scattered light from Earth's atmosphere is too weak compared to the direct sunlight of the second source, and thus can be ignored. Therefore, the second source can be used to describe the properties of dust.

In the experiment, the objective lens is imaged after each measurement of the corona. We denote such lens images as $O_0, O_1, O_2$, and $O_3$.

\subsection{Experimental Procedure}
Two experiments were conducted on November 17, 2021, from 9:00 to 12:30 a.m. (local time), when the weather was clear and cloudless. The experimental procedure is the following:
\begin{enumerate}
	\item Wait for dust to accumulate on the objective lens, approximately equivalent to not wiping for 4 -- 7 days.
	\item Image the corona and objective lens and measure the solar intensity, recorded as $C_0$, $O_0$, and $s_0$.
	\item Slightly clean the objective lens and repeat step ii to obtain $C_1$, $O_1$, and $s_1$.
	\item Slightly clean the objective lens again and repeat step ii to obtain $C_2$, $O_2$, and $s_2$.
	\item Completely clean the objective lens and repeat step ii to obtain $C_3$, $O_3$, and $s_3$.
\end{enumerate}

The solar-radiation intensity is obtained by a calibrated commercial weather station-Davis Vantage Pro2 of the Astronomical Site Monitoring System (ASMS, \citealp{GMG}) at Lijiang Observatory. It measures the solar radiation-intensity at a resolution of $1\rm \ W/m^2$ and an accuracy of 5\%.
\section{Data Analysis}
\label{data_analysis}
\begin{figure}
	\centering
	\includegraphics[width=\linewidth]{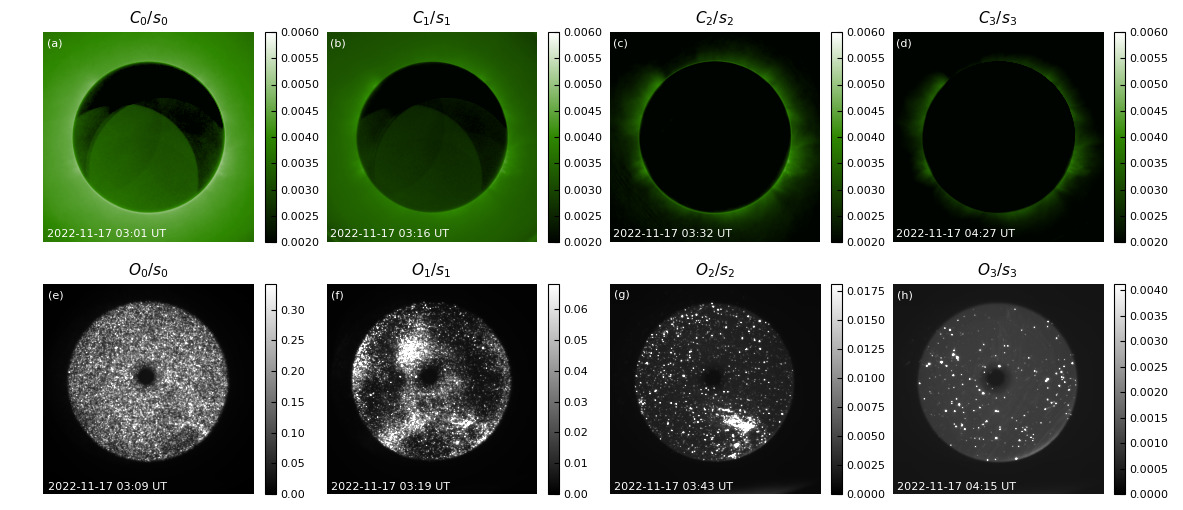}
	\caption{A set of experimental data. The first row shows the corona image (log scale), while the second row displays the objective image. Both images have been adjusted for solar-radiation intensity. From left to right, the amount of dust on the objective decreases sequentially, leading to a gradual reduction in the scattering background in the first-row image and a gradual decrease in the number of scattering points in the second-row image.	}
	\label{data0}
\end{figure}

\begin{figure}
	\centering
	\includegraphics[width=\linewidth]{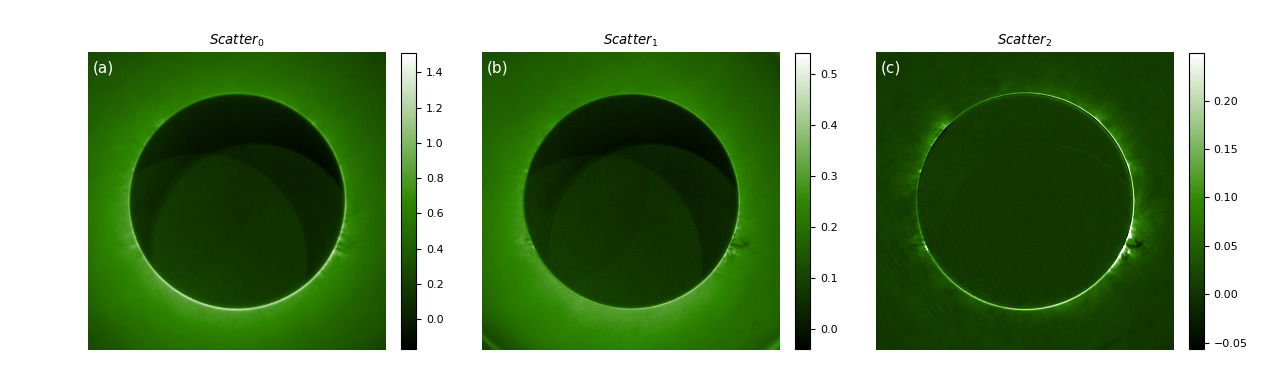}
	\caption{The scattering background of Figures~\ref{data0}a-c obtained using Equation~\ref{c/s}. The intensity gradually weakens from left to right. }
	\label{back}
\end{figure}

The data resulting from one of the two experiments is shown in Figure~\ref{data0}. 
From left to right, the dust on the objective gradually decreases, along with the number of scattering points in the lens images. Note that, the scattering background intensity from Figures~\ref{data0}a-c decreases and there is almost no difference between $C_2/s_2$ and $C_3/s_3$. Three scattering backgrounds can be obtained using Equation~\ref{c/s}, as shown in Figure~\ref{back}. 

It is worth mentioning that some coronal structure can still be seen in the scattering background given in Figure~\ref{back}c. The background given by Equation~\ref{c/s} contains two parts, one is the true scattering background and the other is the coronal difference produced by the constantly changing solar corona observed from the ground. Therefore, the time interval between the different images of a single experiment should be short enough to reduce the effect of the coronal variations. The longest time interval in our experiments is about one hour, which has been shown to have almost no influence on the analysis of the scattering background (\citealp{sha23}). 

In the following, two models will be used to describe the data, which depend on two and three parameters, respectively. The former does not consider the angular distribution of the scattering points, while the latter does.
\subsection{Two-Parameter Model}
\label{2pm}

The radial profiles of the scattering background (the median at each radius) are shown in Figure~\ref{median} (left), along with the linear fitting performed on them, showing strong correlations. Therefore, the slopes ($k$) and intercepts ($b$) of these lines can be used as the parameters of the scattering background and they are related to the corresponding dust. 

The dust is parameterized by the total intensity of scattering points in Figures~\ref{data0}e and f. Using Otsu's segmentation method (\citealp{otsu}), a threshold is calculated for each lens image. Areas with pixel values greater than this threshold are considered dust-scattering points and the intensity of each point can be obtained by summing the pixel values of each area.

Assuming a spatially uniform distribution of scattering points, only one parameter, the total intensity of the scattering points ($I$), can be used  to parameterize the dust. Therefore, the parameters of the scattering background, $k$ and $b$, are considered functions of $I$. As shown in Figure~\ref{median}, all five $k$ and $b$ obtained from the two experiments (one measurement was discarded due to excessive dust) are linearly fitted to $I$, obtaining:

\begin{figure}
	\centering
	\includegraphics[width=\linewidth]{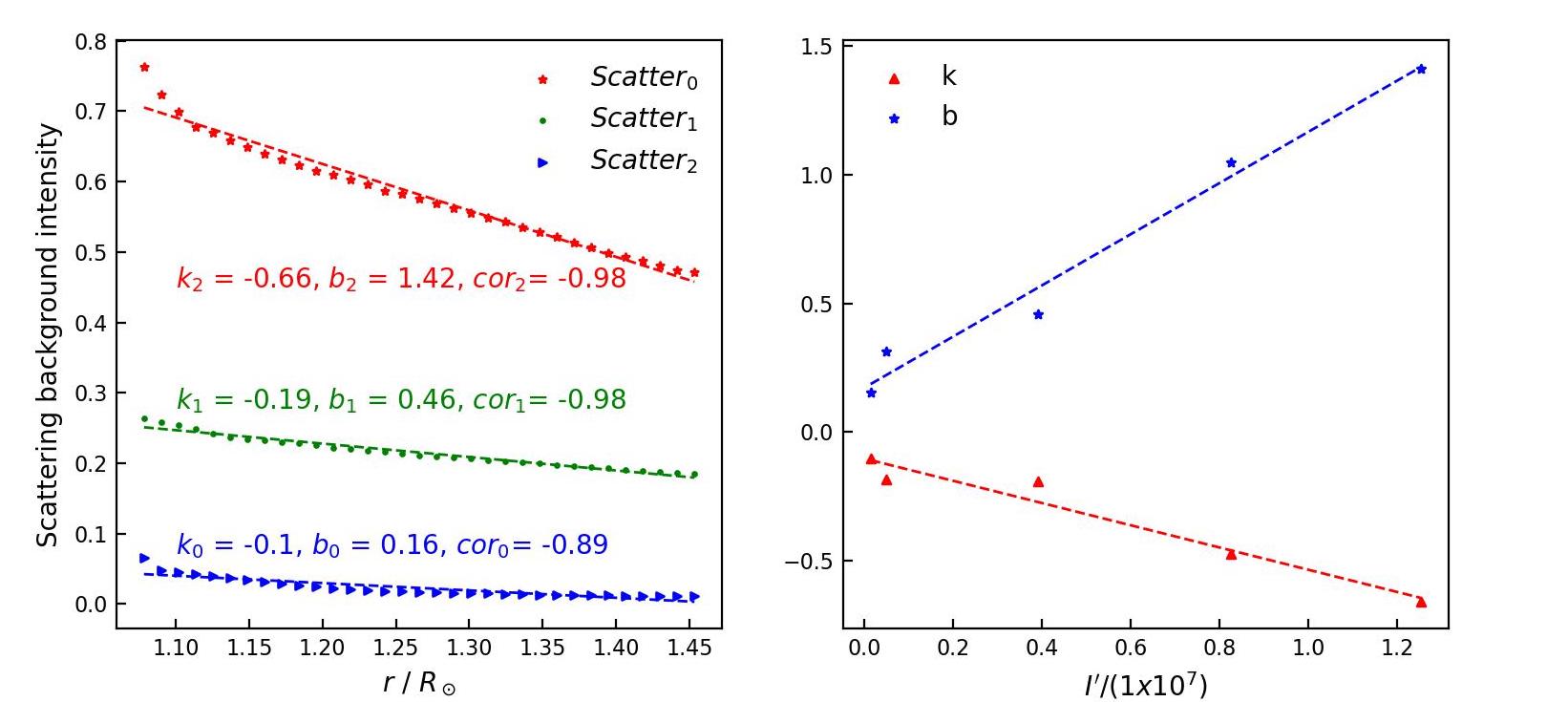}
	\caption{Left: The radial distribution of the scattering background of Figure~\ref{back}, where $k_i$, $b_i$, and $cor_i$ correspond to the slope, intercept, and Pearson correlation coefficients of the least squares fitting. The $y$-axis represents the median intensity of the scattering background obtained from Equation~\ref{c/s}. Right: The relationship between $k, b$ calculated in the left figure and $I'$, where $I'=I/s$. }
	\label{median}
\end{figure}
\begin{equation}
	\label{kibi}
	k(I')=-0.43\times 10^{-7}I'-0.1\ \ \ \ b(I')=0.99\times 10^{-7}I'+0.17,
\end{equation}
where $I'=I/s$. Using Equation~\ref{kibi}, the scattering background can be parameterized with the heliocentric distance ($r$) and the total intensity of the scattering points I, as follows:

\begin{equation}
	\label{eq_2pm}
	\begin{array}{lr}

	\mbox{Scatter}(r,I)=k(I')\times  r/\mathrm{R_\odot}+b(I')	\\
		\ \ \ \ \ \ \ \ \ \ \ \ \ \ \ \ \ =(0.99-0.43\ r/\mathrm{R_\odot})\times 10^{-7}\ I/s+0.17-0.1\ r / \mathrm{R_\odot}.
	\end{array}
\end{equation}

A simulated background image (Scatter$_{2D}$) can be obtained, and used to correct the coronal image as follows:
\begin{equation}
	\label{correct}
	\mbox{Corona}'=\mbox{Corona}-s\times \mbox{Scatter}_{2D}.
\end{equation}

\begin{figure}
	\centering
	\includegraphics[width=\linewidth]{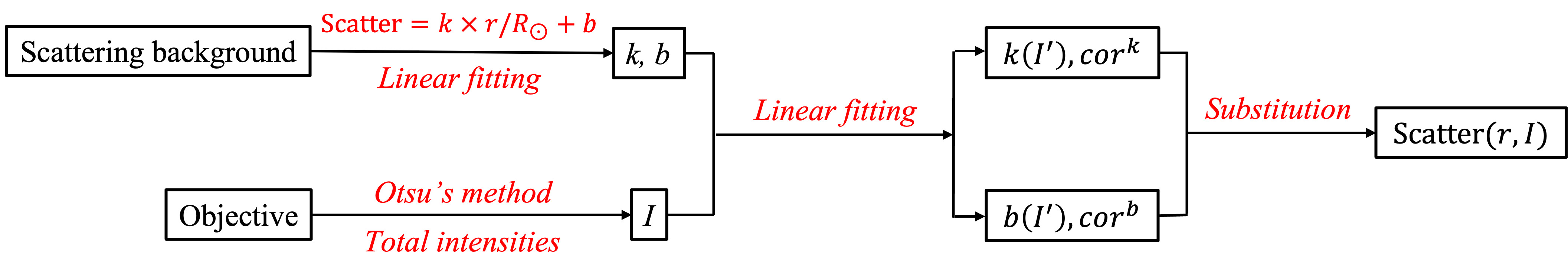}
	\caption{The flowchart of the calibration process. The inputs are the images of the scattering background and the objective. The radial distribution of the scattering background is linearly fitted to obtain $k$ and $b$, satisfying $\mbox{Scatter}=k\times r/R_\odot+b$. Otsu's threshold segmentation method is applied to the objective images to obtain the total intensities of the scattering points $I$. Then, these parameters are linearly fitted to obtain $k(I')$, $b(I')$, and their correlation coefficients ($cor^k$ and $cor^b$). Finally, the two distributions are substituted back to the original equation to obtain $\mbox{Scatter}(r,I)$.}
	\label{flowchart}
\end{figure}

The above calibration process is summarized by the flowchart in Figure~\ref{flowchart}. The inputs to the flowchart are the scattering background and objective image. After a series of parameterization and data fitting, Equation~\ref{eq_2pm} is obtained.

\subsection{Three-Parameter Model}
The angle of the scattering background is not included in the two-parameter model, because we assumed a uniform dust distribution. Therefore, the model will not be accurate when the scattering points are not uniformly distributed.

\begin{figure}
	\centering
	\includegraphics[width=\linewidth]{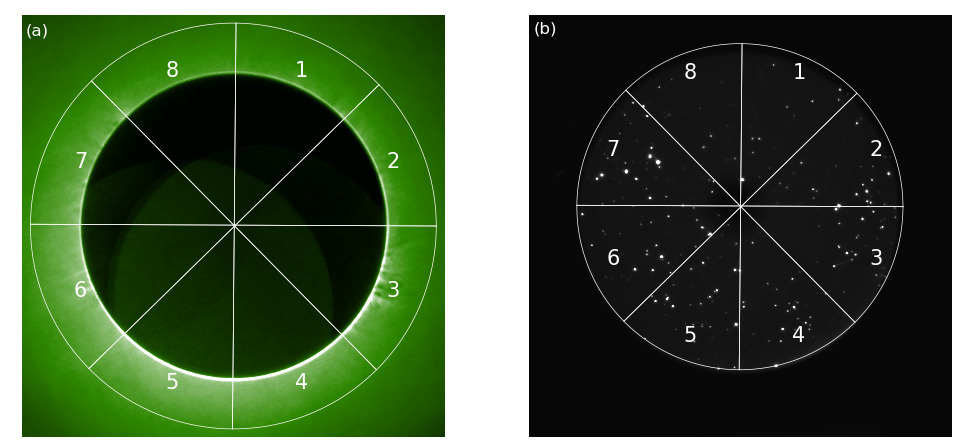}
	\caption{The left image illustrates the scattering background, whereas the right image shows the objective lens. Each figure is divided into $N$ circular sectors by solid white lines (here $N=8$), and each region is labeled from 1 to $N$.
	}
	\label{sector}
\end{figure}

To find the relationship between the angular distribution of the scattering points and the scattering background, both the objective images and the scattering-background images are divided into $N$ equal circular sectors (Figure~\ref{sector}). 
We randomly select the $m$th and $n$th ($1\le m,n\le N$) sectors from the objective images and the scattering-background images respectively, and use them as the inputs to the flowchart in Figure~\ref{flowchart}. Thus, we obtain $k(I')$, $b(I')$, and their linear correlation coefficients, denoted as $cor^k_{m,n}, cor^b_{m,n}$.
By iterating through all values of $m$ and $n$, a total of $2N^2$ correlation coefficients can be obtained. 

\begin{figure}
	\centering
	\includegraphics[width=.6\linewidth]{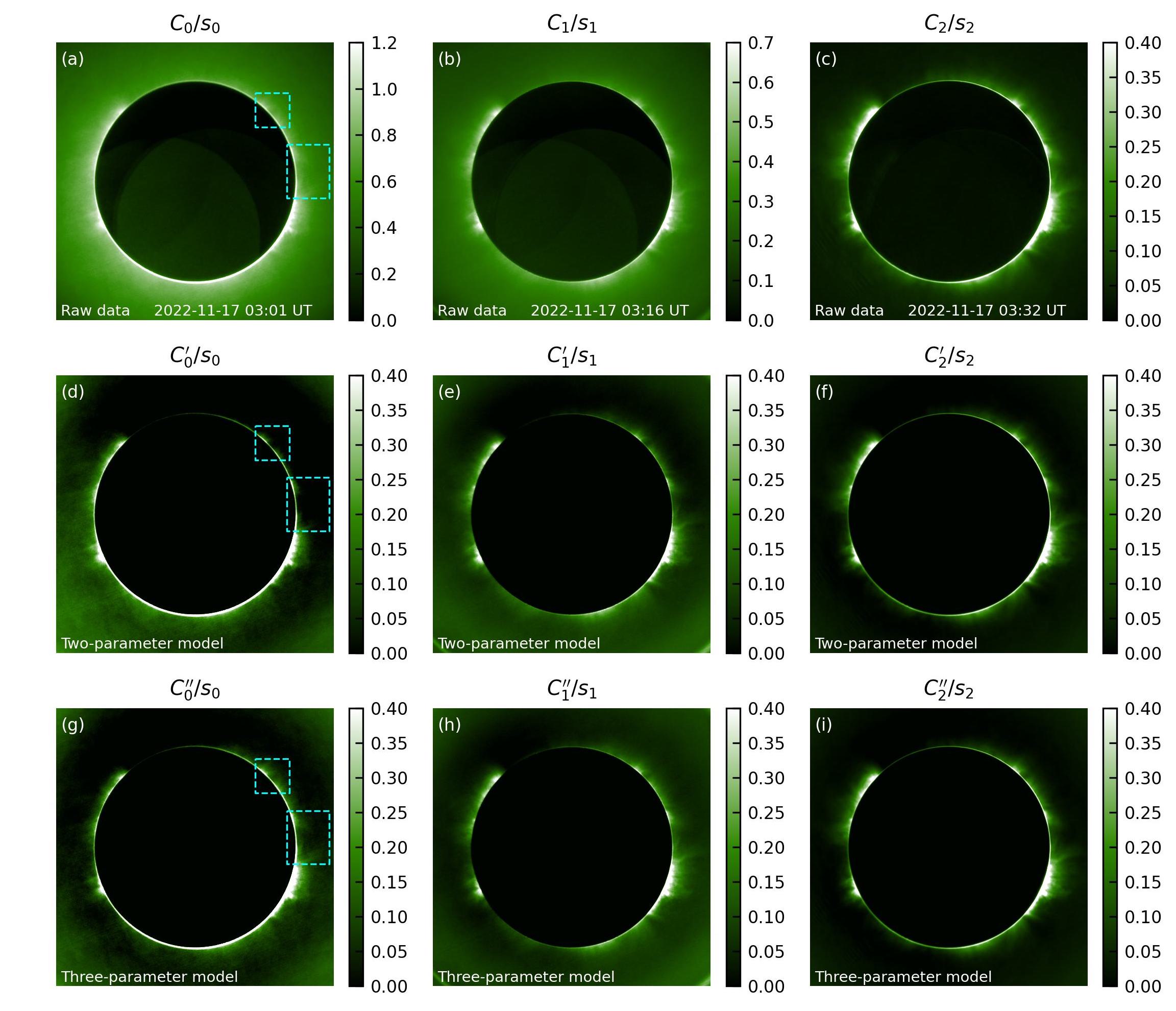}
	\caption{The distribution of average correlation coefficient between the total intensity of the scattering points in the circular sectors and the parameters $k$ (blue), $b$ (orange) of the scattering background versus the distance of the two sectors involved. $N$ is set to 360.}
	\label{cor}
\end{figure}

The distance between the two circular sectors is defined as follows:
\begin{equation}
	\mbox{Distance}=	
	\left\{
	\begin{array}{lr}
		n-m & m\le n \\
		n-m+N &m>n
	\end{array}
	\right. .
	\label{eq-dis}
\end{equation}
We compute the average correlation coefficient versus the distance of all circular sectors pairs with N=360, i.e. the distance is equal to the difference of the central position angles of the two circular sectors involved ($\phi$), see Figure~\ref{cor}.
It can be seen that the largest correlation occurs when the distance between the two circular sectors is approximately $180^\circ$. The weakest correlation corresponds to a distance of about zero ($\phi \approx 0$). This antisymmetric correlation is attributed to the disparity in imaging directions between the objective and the corona. We conclude that the scattering background in any given direction $\theta$ is contributed to varying degrees by the intensity of scattering points in all angular directions. 

A sinusoidal function is fitted to the curves in Figure~\ref{cor} to determine an ideal maximum value at $\phi=191^\circ$. We then modify the parameter $I$ in Equation~\ref{eq_2pm} as follows:
\begin{equation}
	I_\theta(\theta)=I-\sum_{\phi=1}^{360}I_{\theta+\phi}\times \sin(\theta+\phi+79^\circ),
\end{equation}
where $I_{\theta+\phi}$ represents the total intensity of scattering points in the $(\theta+\phi)$th circular sector. Then, Equation~\ref{eq_2pm} can be rewritten as:
\begin{equation} 
	\label{eq_3pm}
	\mbox{Scatter}(r,I_\theta(\theta))=(0.99-0.43\ r/\mathrm{R_\odot})\times 10^{-7}I_\theta(\theta)/s+0.17-0.1\ r/\mathrm{R_\odot}.
\end{equation}

\section{Results}
\subsection{Model Validation}
\begin{figure}
	\centering
	\includegraphics[width=\linewidth]{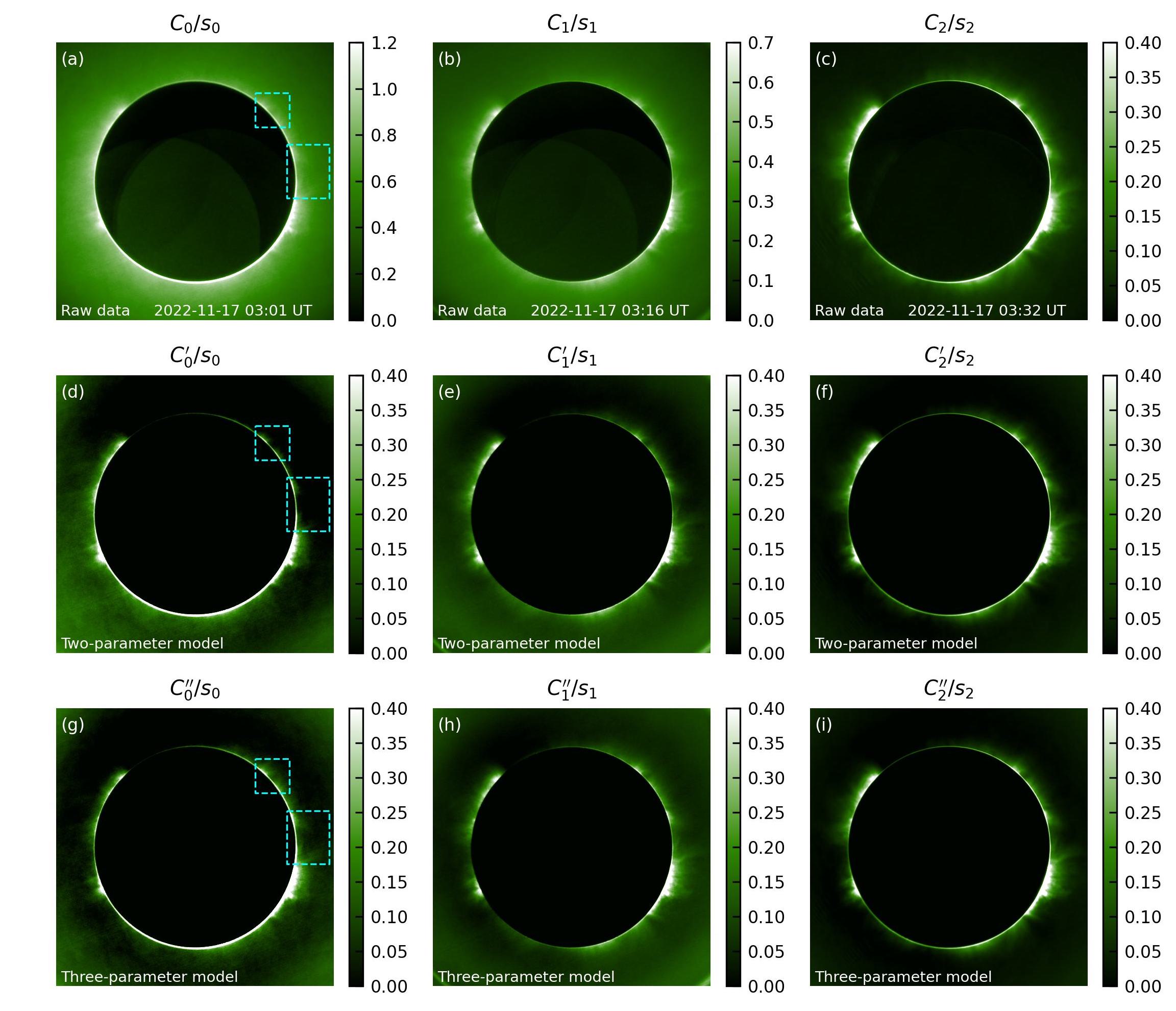}
	\caption{Corrected coronal data. Panels of the first row show the original data, the same as Figures~\ref{data0}(a)$\sim$(c). Panels in the second and third rows represent the images corrected using the two- and three-parameter models, respectively. Note that, for the structures enclosed by the dashed boxes, panel g exhibits a superior detail structure compared to panel d, indicating a more effective correction by the three-parameter model. This is more evident in Figure~\ref{lg}.
	}
	\label{33fig}
\end{figure}
The two models described in Section~\ref{data_analysis} were separately applied to correct the coronal images, as shown in Figure~\ref{33fig}.
The pixel values of the coronal images processed by both models fall within a similar range, suggesting the effectiveness of the models in removing an important proportion of the superimposed dust-scattering background. Additionally, the second model outperforms in capturing finer details, compare Figures~\ref{33fig}a, d, and g, whereas the remaining sets of corrected images exhibit no significant difference. This indicates that the advantage of the three-parameter model becomes more prominent under high dust concentrations.

\begin{figure}
	\centering
		\includegraphics[width=\linewidth]{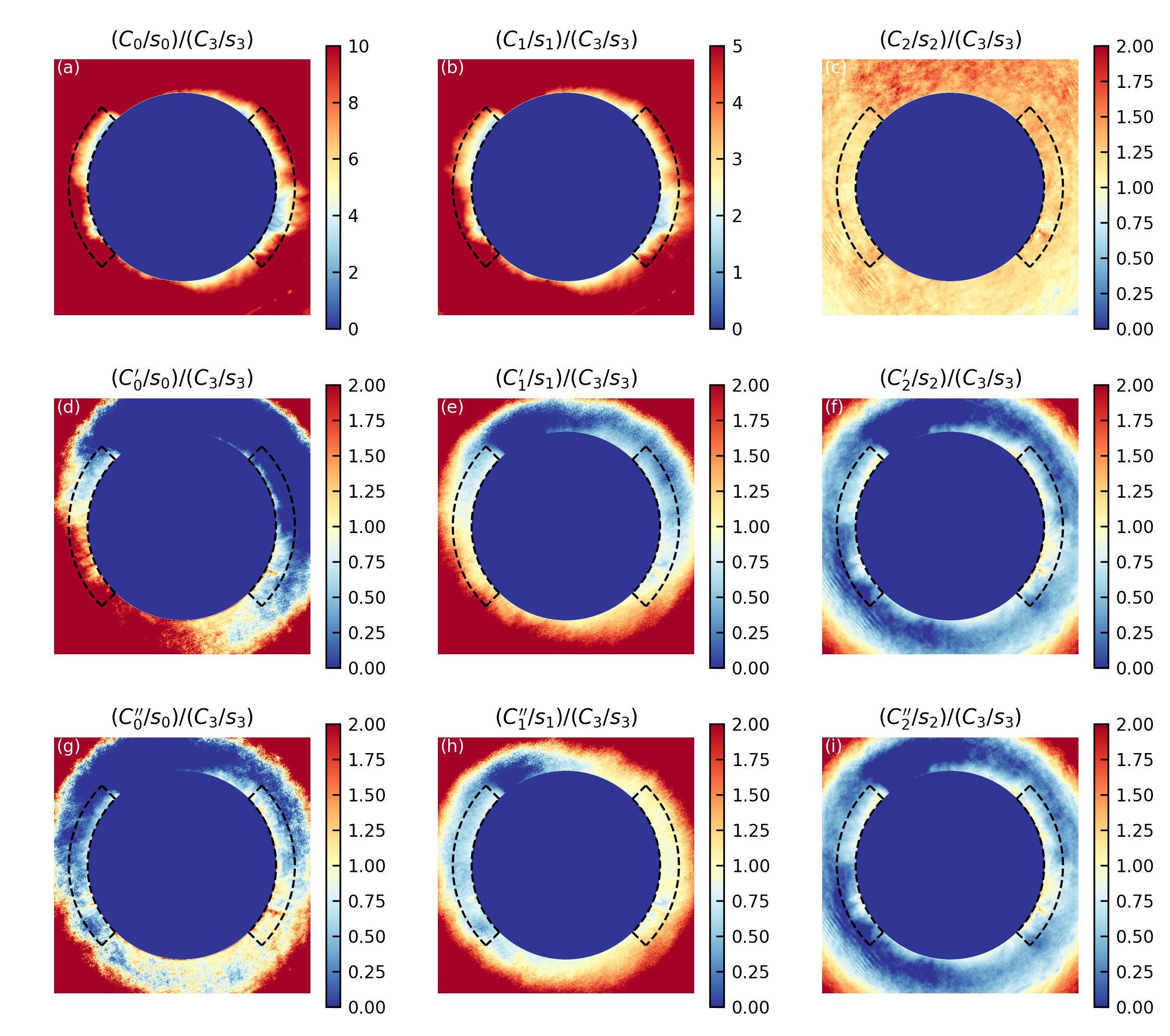}
		\caption{Ratio of the raw and corrected coronal image in Figure \ref{33fig} by the quasidust-free coronal image ($C_3/s_3$). The occulter regions are masked. The areas within the frames in panel g are yellow  ($C_0''/s_0\approx C_3/s_3$), while in panels (a) and (d) they are red ($C_0/s_0 \gg C_3/s_3$) and  blue ($C_0'/s_0 \ll C_3/s_3$), respectively. 
		The regions enclosed by the black dashed line are used to calculate the mean relative error in Figure \ref{accuracy}.}
		\label{lg}
\end{figure}

More clearly, we divide the corrected image by the cleanest coronal image ($C_3/s_3$) 
, as shown in Figure~\ref{lg}. 
The images of the first row represents the error of the raw data, with most of the areas in red indicating a large amount of scattering background in the raw data. The following two rows represent the results of the correction using two models, with the original red area turning yellow or blue, indicating that the correction effectively removes the scattering background and improves the signal-to-noise ratio.

When there is a large amount of dust, a large blue area appears in the corrected image, which is relatively small in the three-parameter model corrected images. However, the colors in Figures~\ref{lg}f and i are slightly bluish, indicating an overcorrection. This is due to the fact that $O_3$ is not completely clean in the experiment, which increases the error of the model when there is very little dust. In future experiments, we will find ways to improve the cleanliness of $O_3$ and improve the accuracy of the model.

\begin{figure}
	\centering
	\includegraphics[width=\linewidth]{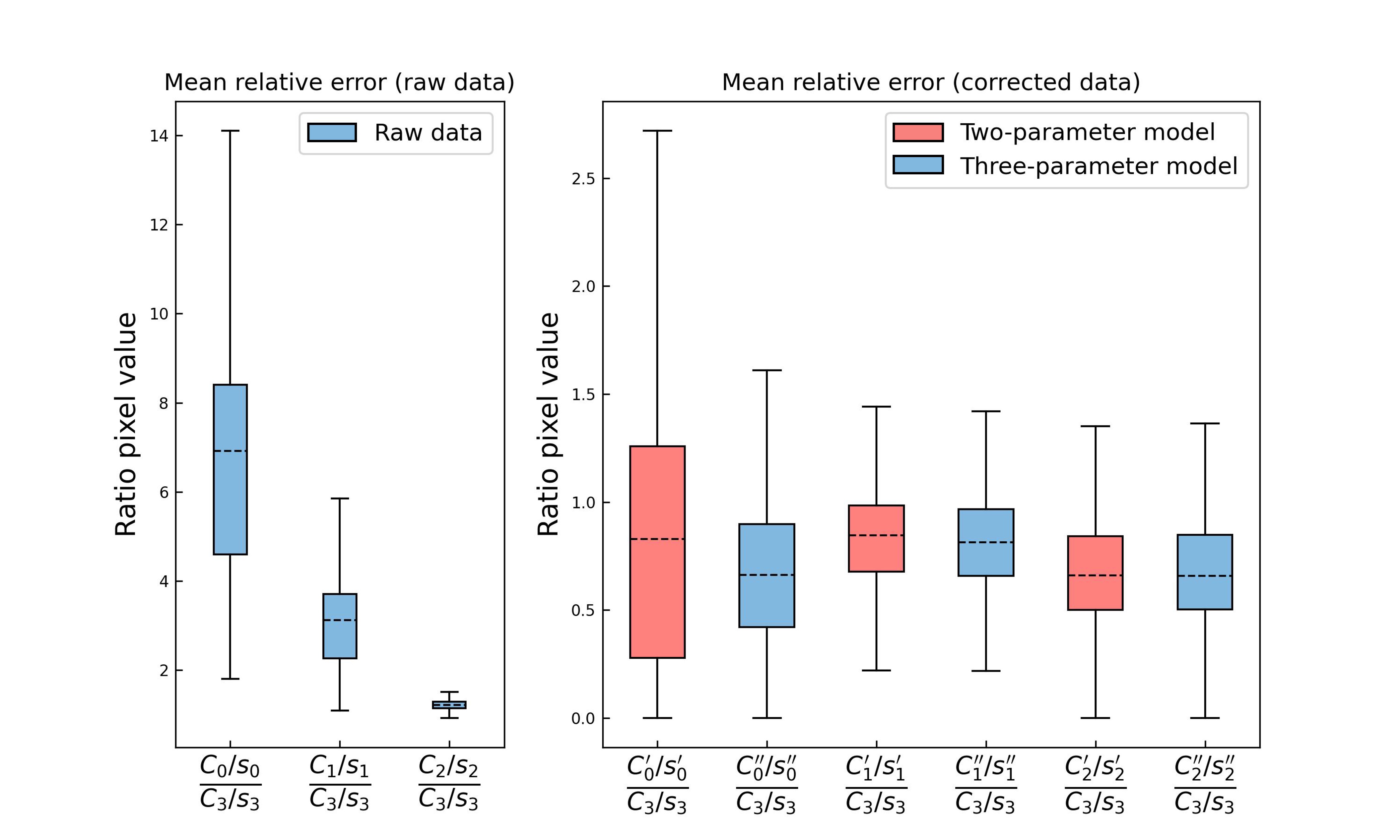}
	\caption{The boxplot of the relative error with respect to $C_3/s_3$. It is calculated from the pixel values of the areas enclosed by the black dashed line in Figure~\ref{lg}. The line in the middle of the box represents the average value, and the top and bottom sides of the box represent the 25th and 75th percentile values. The left panel shows the errors of the raw data and the blue and red boxes in the right panel represent the correction results of the two- and three-parameter models, respectively. 
}
	\label{accuracy}
\end{figure}

The differences between the corrected and the quasidust-free ($C_3$) coronal images are quantified by the
mean relative error with respect to $C_3/s_3$ . 
It can be reflected by the pixel values in the low-latitude ($<45^\circ$) region  within $1.2\ \rm R_\odot$ of Figure~\ref{lg} (the regions enclosed by the black dashed line in Figure~\ref{lg}).
We show the mean relative error with respect to $C_3/s_3$, see Figure~\ref{accuracy}. The average value after correction is greatly reduced and close to 1 but slightly less than 1, indicating overcorrection. When there is a large amount of dust, e.g., $C_0$, the three-parameter model can better reduce the variance of the correction results. When the amount of dust is small, there is no significant difference in the results between the two models. It should be noted that this value also includes any error produced by intrinsic changes in the solar corona and by estimating $s_i$. 

\begin{figure}
	\centering
	\includegraphics[width=\linewidth]{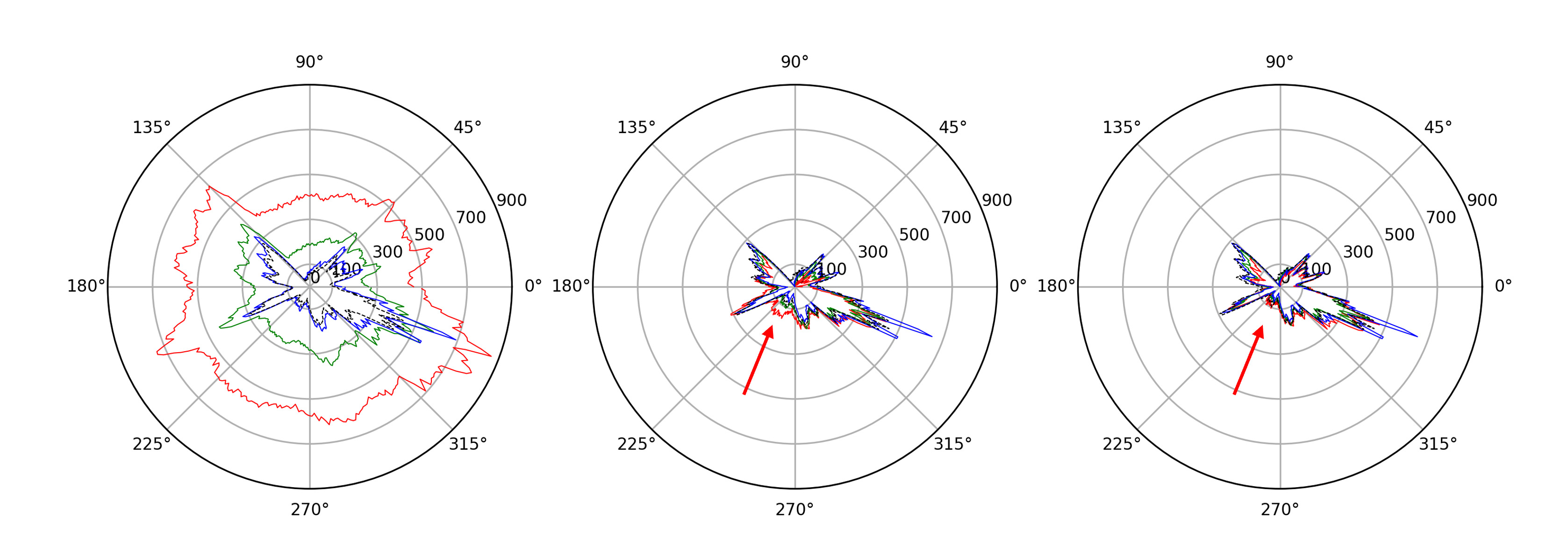}
	\caption{Angular intensity distribution at $r = 1.08\ \rm R_\odot$ in the coronal images of each row in Figure~\ref{33fig}. The dashed line represents the corona obtained when the objective lens is at the cleanest ($C_3$), while the red, green, and blue lines correspond to $C_0$, $C_1$, and $C_2$. The leftmost panel represents the original data, while the middle and rightmost panels present the results corrected using the two- and three-parameter models, respectively. In the latter two, the corrected intensity shows a significant reduction and better matches the intensity of the solar corona, e.g., see the red arrows.}
	\label{fig_polar}
\end{figure}

As a more quantitative comparison, the angular variation of intensity at $r=1.08\ \rm R_\odot$ was extracted for each image in Figure~\ref{33fig}, see Figure~\ref{fig_polar}. 
After removing the scattering background, the angular distribution becomes consistent with $C_3/s_3$, which is the cleanest condition. Additionally, they are more consistent when applying the three-parameter model correction, which not only validates the effectiveness of the model but also emphasizes the necessity of using the three-parameter model.

Future observations hold the potential for enhancing data utilization and accuracy by periodically capturing images of the objective lens to obtain $I(\theta)$ and correcting the coronal data using Equation \ref{eq_3pm}.

\subsection{Radial Distribution of Dust}
\begin{figure}
	\centering
	\includegraphics[width=.6\linewidth]{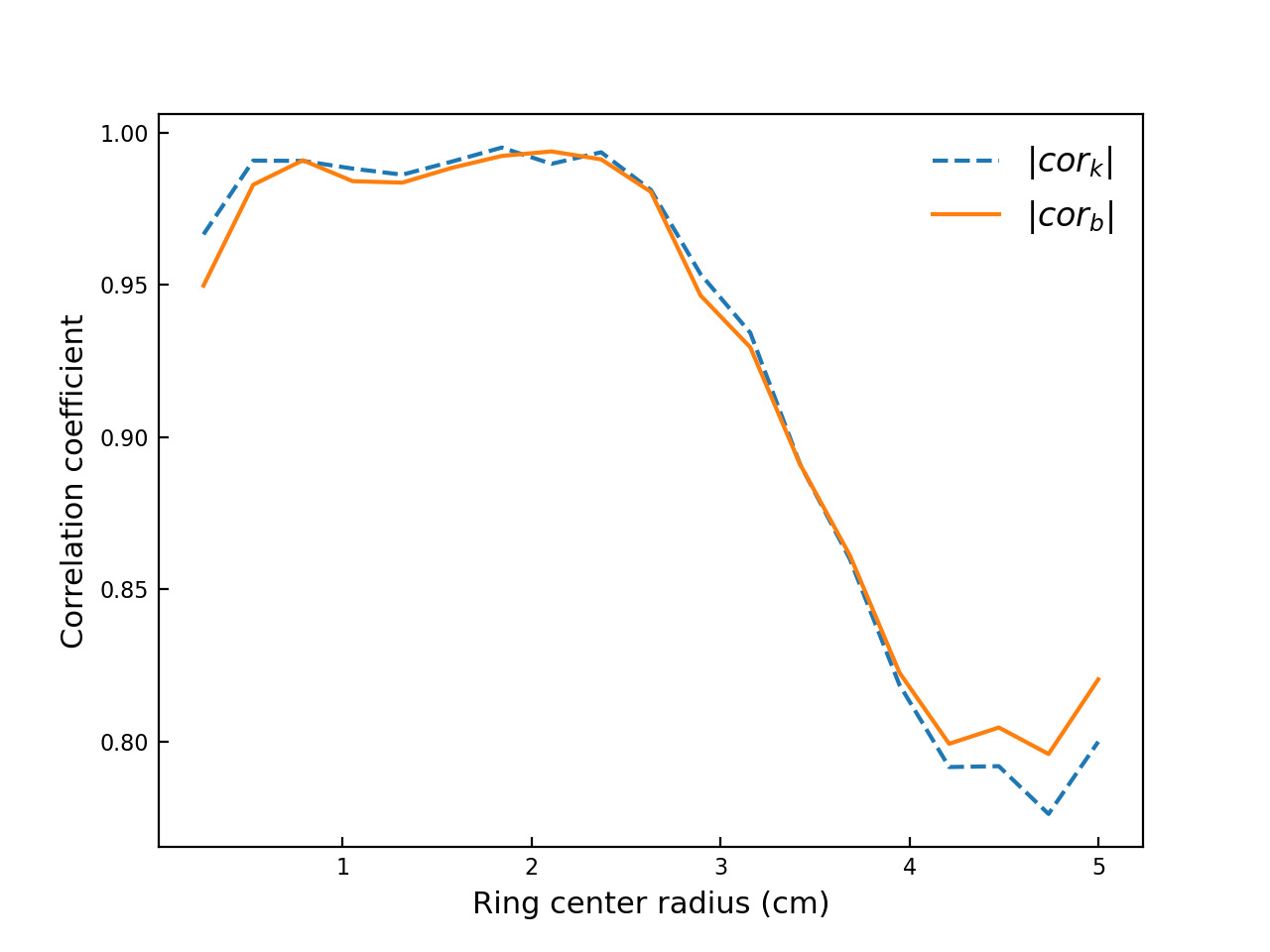}
	\caption{The distribution of the correlation coefficients between the total intensity of the scattering points in concentric rings of different radii and the parameters $k$ and $b$ of the scattering background. Recall that the objective lens diameter is 10 cm.}
	\label{ring}
\end{figure}

Given that different angular distributions of scattering points can result in different scattering backgrounds, it is expected that their different radial distributions will also have varying impacts. Similar to the approach described in Section 3.2, in this case, the objective image is no longer divided into circular sectors, but into concentric rings. 
The intensity of scattering points per unit area in different concentric rings is calculated separately, followed by determining the correlation between these intensities and the parameters of the scattering background, $k$ and $b$, as illustrated in Figure~\ref{ring}.

It can be seen that the correlation between the scattering background and the scattering points, within a radius of approximately 2.5 cm from the center of the objective lens, is very high, approaching a value of 1. Conversely, as the radius of the ring increases beyond about 2.5 cm, the correlation with the scattering background progressively diminishes. 
This suggests that dust particles in the outer region of the objective lens contribute less to the scattering background.
\section{Conclusion}

We evaluate the influence of dust particles on coronagraph objective lenses on coronal imaging and correct the coronal data using empirical models to enhance its quality. YOGIS was used to capture images of dust-induced scattering points and the scattering background on the coronal image. The connection between them was analyzed using two models. The first two-parameter model describes a radially symmetric dust-scattering background based only on the radial position and the intensity obtained from the objective-lens image. The three-parameter model, also takes into account the angular distribution of scattering points, thereby allowing us to establish a connection between the scattering background and height, angle, and intensity of the scattering points (Equation~\ref{eq_3pm}). 

We conclude that it is advantageous to periodically capture objective images to obtain $I(\theta)$ and then correct the coronal data using Equation~\ref{eq_3pm}. 
The quality of the resulting coronal data can be improved to enhance coronal structures, which is beneficial for the discovery of coronal mass ejections (CMEs), low-speed out flows  (e.g. \citealp{Xia03AA,Xia04AA,Tian21SoPh}) , among others.
We consider that this model is an important step in the precise calibration of large ground-based coronagraphs.
Our technique can be particularly useful to support future routine measurements of the coronal magnetic field using coronagraphs (e.g. \citealp{Liu08,liu09,liu10}).

In addition, we note that the dust on the inner part of the objective contributes more to the scattering background, which can serve as a valuable guide for the development of superior scattering-background models in the future.

The findings of this study highlight the importance of dust on the objective lens and its impact on the stray light of YOGIS. The proposed method involves capturing images of the corona and the objective lens, making it a simple and clear process. This technique can be applied to various internally occulted coronagraphs and serves as a crucial reference for accurate calibration of images obtained with other ground-based coronagraphs. This technique may be also applicable to other optical astronomical instruments. It should be noted that the actual calibration images presented in this study are only applicable to YOGIS due to variations in optical systems, filters, cameras, and other factors that differ among coronagraphs. As a result, the conclusions derived from this study cannot be directly extrapolated to other coronagraphs. Nevertheless, conducting experiments using the methodology outlined in this paper can provide calibrations applicable to other instruments. 

The upcoming 20-cm coronagraph of the Chinese Meridian Project Phase II, which will be stationed at the Lijiang Observatory, is also considering adding an objective-lens imaging system to perform simultaneous lens imaging during coronal observations. Furthermore, analyzing the scattered light generated by dust particles on the objective lens can also be used to diagnose the current level of stray light in the instrument.

To improve the quantity and accuracy of data in this experiment, we need to address the limitations in the observing conditions. One way to do this is by minimizing the duration of each set of experiments to reduce the effect of solar rotation and the evolution of the corona. We also plan to conduct experiments at the future coronagraph site in Daocheng, Sichuan province, which is situated at a higher altitude of 4700 m (\citealp{liu18}). This can help to gather more high-quality data and enhance the calibration accuracy.

%

%

%

%
\begin{acks}
We sincerely thank Prof. T. Sakurai-san and the NAOJ team for the 10 cm coronagraph collaboration. This work is supported by the National Natural Science Foundation of China
under grants (12173086, 11873090, 12373063, 11533009), the Light of West China program of CAS and Yunnan Key Laboratory of the Solar Physics and Space Science [202205AG070009], and the National Key R\&D Program of China No. 2022YFF0503800. We acknowledge the data resources from the National Space Science Data Center, National Science \& Technology Infrastructure of China (\href{http://www.nssdc.ac.cn}{http://www.nssdc.ac.cn}).
\end{acks}

\begin{authorcontribution}
Feiyang Sha, Xuefei Zhang, and Tengfei Song conducted the experiments. Feiyang Sha, Yu Liu, and Xuefei Zhang wrote the main manuscript text. Feiyang Sha prepared the figures. All authors reviewed the manuscript. Yu Liu supervised this work and Xuefei Zhang partly assisted the assembly and guidance of the instruments in Lijiang Observatory.
\end{authorcontribution}

%
 \begin{dataavailability}
The data that support the findings of this study are available from the authors, upon reasonable request.
 \end{dataavailability}

%

\begin{conflict}
We declare that we have no conflict of interest.
 \end{conflict}

%
%
\bibliographystyle{spr-mp-sola}
\bibliography{template} 
 
%
%
%
%

\end{article} 
\end{document}